%% file: DiscreteKL_v2.tex
\documentclass[useAMS,usenatbib,referee]{biom}

\input{preamble_ENAR2022.tex}

\usepackage[figuresright]{rotating}
\def\bSig\mathbf{\Sigma}

\title[Kullback-Leibler-Based Discrete Failure Time  Models]{Kullback-Leibler-Based Discrete Failure Time Models for Integration of
Published Prediction Models with New Time-To-Event Dataset}

\author{Di Wang$^{1}$, 
Wen Ye$^{1}$, Randall Sung$^{2}$, Hui Jiang$^{1}$, Jeremy M.G. Taylor$^{1}$, Lisa Ly$^{3}$, and Kevin He$^{1,*}$\email{kevinhe@umich.edu} \\
$^{1}$Department of Biostatistics, University of Michigan, Ann Arbor, Michigan, U.S.A. \\
$^{2}$Section of Transplantation Surgery, University of Michigan, Ann Arbor, Michigan, U.S.A.\\
$^{3}$ Department of Urology, Temple University, Philadelphia, Pennsylvania, U.S.A.}

\begin{document}





\pagerange{\pageref{firstpage}--\pageref{lastpage}} 




\label{firstpage}


\begin{abstract}
Prediction of time-to-event data often suffers from rare event rates, small sample sizes, high dimensionality and low signal-to-noise ratios.
 Incorporating published prediction models from large-scale studies is expected to improve the performance of prognosis prediction on internal individual-level time-to-event data. However, existing integration approaches typically assume that underlying distributions from the external and internal data sources are similar, which is often invalid. 
To account for challenges including heterogeneity, data sharing, and privacy constraints, we propose a 
discrete failure time modeling procedure, which utilizes a discrete hazard-based Kullback-Leibler discriminatory information measuring the discrepancy between the published models and the internal dataset. Simulations show the advantage of the proposed method compared with those solely based on the internal data  or published models. We apply the proposed method to improve prediction performance on a
kidney transplant dataset from a local hospital by integrating this small-scale dataset with published survival models obtained from the national transplant registry.
\end{abstract}

%



\maketitle


%

\section{Introduction}
\label{s:intro}

Prognosis prediction is an important topic in survival analysis. Exploration of newly collected biomarkers from recent studies is critical to improve prediction performance and facilitate clinical practice. However, information from an individual study is usually inadequate to make stable predictions, and the resulting prediction models often suffer from rare event rates, small sample sizes, high dimensionality, and low signal-to-noise ratios.
Incorporating external information from large-scale studies, such as prediction models built upon national disease registries, is expected to improve prediction performance \citep{liu2014}. 

Our endeavor is motivated by the study of end-stage renal disease (ESRD), which represents 7.1\% of the entire Medicare budget in the United States \citep{johansen22}. While a kidney transplant is the preferred treatment for ESRD \citep{Wolfe1999}, the demand far exceeds the supply, with only 3.9\%  of incident ESRD patients having received a kidney transplant in 2019. To optimize treatment strategies for ESRD patients, an important aspect is to increase organ procurement efficiency and manage post-transplant graft failure risk. For these purposes, the Scientific Registry of Transplant Recipients (SRTR) maintains risk adjustment models for post-transplant graft survival \citep{Snyder2016}. However, it is widely recognized that SRTR models only include conventional predictors captured in the national registry, and that the prediction performance of SRTR models is limited \citep{kasiske2019}. Hence, the assessment of additional risk factors is necessary to improve the risk adjustment by SRTR. Due to technical and financial restrictions, pilot programs to explore new biomarkers often have small sample sizes and short follow-up time, which can lead to unstable prediction and poor calibration. Thus, researchers are considering integrating external information from published prediction models, such as the aforementioned SRTR models, with newly collected data to facilitate the risk adjustment.

Existing integration methods were typically developed under a strong assumption that the external and internal studies are homogeneous; i.e. 
the internal data is collected from the same underlying probability distribution from which the external summary-level information is derived.
For example, empirical-likelihood-based methods have been widely used to combine auxiliary information and parametric likelihood \citep{chen1993,qin2000,Chaudhuri2008,Qin2015}. More recently, \citet{chatterjee16} developed a constrained semiparametric maximum likelihood estimation procedure using parameters from an external model to improve the estimation efficiency. In the context of survival analysis, under the assumption that the external auxiliary information and internal data are homogeneous, \citet{huang2016} developed a double empirical likelihood approach based on the Cox model. However, in practice, the  homogeneity assumption is often violated. For instance, the covariate spaces; the data collecting, processing and analyzing procedures; the true underlying generating mechanisms; and context-specific confounders may differ substantially across studies. Ignoring heterogeneity across populations can yield substantially biased parameter estimates and a loss of efficiency \citep{estes17}. 

To relax the homogeneity assumption, \citet{chen2021} and \citet{sheng2021:2} proposed penalized empirical likelihood approaches by introducing a penalty term on the potential discrepancy between the internal cohort and the external summary information. While successful, these methods only apply to external subgroup survival rates at certain time points defined based on a few categorical variables (e.g. 5-year survival probabilities for patients older than age 65),
which precludes the application to the kidney transplant study of the motivating example. 
Specifically, the published SRTR 1-year post-transplant graft survival model is derived from a Cox proportional hazards model with a total of 59 covariates. Both model coefficients and baseline cumulative hazards during follow-up after the kidney transplant are reported, which enable the prediction of weekly post-transplant survival probabilities. Our goal is to integrate such weekly predicted survival probabilities with a newly collected individual-level time-to-event dataset to improve the post-transplant risk adjustment and prediction.

Our proposed method builds on Kullback-Leibler (KL) discrimination information \citep{KL}, which is a measure of discrepancy between two probability distributions. 
Since the full probability distribution for censored observations in time-to-event data is not observed, one major challenge in developing KL-based integrating methods for time-to-event data is how to accommodate censored information. To fill this gap, we propose a novel discrete hazard KL discrimination information to measure the discrepancy between the published survival prediction model (denoted as external model) and the internal time-to-event data, incorporating both failed and censored observations. A KL-based discrete failure time modeling procedure is developed to improve the prediction performance of the aggregated model.
The proposed method is flexible and general in the following aspects: (1) it integrates a sequence of external survival probabilities defined at multiple time points with the internal individual-level data; 
(2) it is robust to mis-specified external models and incorporates various types of external prediction models, including Cox models, discrete failure time models, and machine learning algorithms, as long as the external model can provide predicted hazard functions; (3) it can simultaneously incorporate multiple external models, identifying the most compatible external information source and diminishing the weights of less relevant ones;  and (4) it is computationally efficient for high-dimensional problems and can be easily extended to machine learning methods, such as gradient boosting.


The rest of the paper is organized as follows: In Section~\ref{s:notation}, we introduce the prerequisite notations. In Section~\ref{s:method}, we propose the discrete hazard KL information for discrete failure time models and introduce the proposed integration framework.  Section~\ref{s:sim} conducts simulations to evaluate the proposed method. In Section~\ref{s:real}, a data analysis of the kidney post-transplant graft survival prediction is presented. Discussions are summarized in Section~\ref{s:discuss}.

\section{Notation and Model Setting}
\label{s:notation}
In this section, we introduce the notations of the discrete failure time model for the internal individual-level data (Section~\ref{sub:internal_model}), and demonstrate that various types of external models can provide a sequence of predicted hazards across discrete failure times (Section~\ref{sub:external_model}). Our goal is to integrate such external predicted hazards with the internal individual-level data to enhance prediction performance.
\subsection{Internal Discrete Failure Time Model}
\label{sub:internal_model}
In clinical studies, event times are often measured on a discrete scale, which motivates the use of a discrete failure time model \citep{Kalbfleisch2002, Tutz16}. Let $T_i$ denote the event time of interest and $C_i$ be the censoring time for subject $i$ in
the internal dataset, $i=1,\ldots, n$, where $n$ is the sample size of the internal cohort. Let $t_1, \ldots, t_K$ be the distinct failure times indexed by $k=1,  \ldots, K$. 
Let $\mathcal{D}_{k}$ and $\mathcal{C}_{k}$ denote the set of labels associated with individuals failing and censoring at time $t_k$, respectively. Let $\textbf{Z}_{i}$ be a possibly time-dependent covariate vector for the $i$-th subject. Assume that $T_i$ and $C_i$ are independent, given the value of $\textbf{Z}_{i}$. Let $\lambda(t_k;\textbf{Z}_{i})=P(T_i=t_k|T_i\geq t_k, \textbf{Z}_{i})$ be the hazard function at time $t_k$ for the $i$-th subject with covariate $\textbf{Z}_{i}$. The  likelihood function is given by
\begin{equation}
   L=\prod_{k=1}^{K}\left\{\prod_{i \in \mathcal{D}_{k}} f(t_k;\textbf{Z}_{i})  \prod_{i \in \mathcal{C}_k} S(t_k;\textbf{Z}_{i})\right\},
   \label{eq:likelihood}
\end{equation}
where
\begin{equation}
   S(t_k;\textbf{Z}_{i})=P(T_i>t_k|\textbf{Z}_{i})= \prod_{\ell=1}^k \{1-\lambda(t_{\ell};\textbf{Z}_{i})\}
 \label{eq:survival}  
\end{equation}
is the survival function, and
\begin{equation}
f(t_k;\textbf{Z}_{i})=P(T_i=t_k|\textbf{Z}_{i})= \lambda(t_{k};\textbf{Z}_{i})\prod_{\ell=1}^{k-1} \{1-\lambda(t_{\ell};\textbf{Z}_{i})\}
 \label{eq:density}  
\end{equation}
is the density function of subject $i$ at time $t_k$.  Let $h:[0,1] \rightarrow [-\infty, \infty]$ denote a monotonically increasing and twice differentiable link function. Consider a general formulation of the hazard $ \lambda(t_k;\textbf{Z}_{i}) = g(\gamma_k + \textbf{Z}^\top_{i} \boldsymbol\beta) $, where $g$ is the inverse function of $h$, $\gamma_k$ is the $h$-transformed baseline hazard at $t_k$, and $\boldsymbol\beta$ is a coefficient vector associated with $\textbf{Z}_{i}$. The log-likelihood is given by
\begin{equation}
 \ell(\bm{\theta})
 = \sum_{i=1}^n\sum_{k=1}^{K}   Y_i(t_k)\left[ \delta_i(t_k) \log\left\{\frac{g(\gamma_{k}  + \textbf{Z}^\top_{i} \boldsymbol\beta)}{1-g(\gamma_{k}  + \textbf{Z}^\top_{i} \boldsymbol\beta)}\right\} 
 + \log\{1-g(\gamma_{k}  + \textbf{Z}^\top_{i} \boldsymbol\beta)\} \right],
 \label{eq:loglik}
\end{equation}
where $\bm{\theta}=(\bm{\gamma}^\top, \bm{\beta}^\top)^\top$, $\bm{\gamma}=(\gamma_{1}, \dots, \gamma_{K})^\top$,   $Y_i(t_k)=I(\min\{T_i,C_i\} \geq t_k)$ is the at-risk indicator and $\delta_i(t_k)=I(T_i=t_k)$ is the failure indicator at time $t_k$.

\subsection{External Predicted Hazards}
\label{sub:external_model}
Suppose we have an external prediction model that takes the input covariate vector $\textbf{X}_{i}$ and returns predicted hazards
$\tilde\lambda(t_k;\textbf{X}_{i})=\tilde{P}(T_i=t_k|T_i\geq t_k, \textbf{X}_{i})$ 
for each subject  in the internal data at each time point, where
$\tilde{P}$ is a failure time distribution corresponding to the external model. To allow the internal data to contain more variables than used in the external prediction models, assume that the values of $\textbf{X}_{i}$ can be determined by the values of $\textbf{Z}_{i}$; for instance, $\textbf{X}_{i} \in \mathcal{X}$,  $\textbf{Z}_{i} \in \mathcal{Z}$ and $\mathcal{X} \subset \mathcal{Z}$.
Two examples of external prediction models are given as follows:

\textit{Example 1 (Predicted hazards based on Cox models).}
In the field of survival analysis, the most commonly used models for predicting survival risks are the Cox models.
For example, in the aforementioned kidney transplant study, the SRTR risk adjustment model for the 1-year post-transplant outcome was derived from a Cox proportional hazards model \citep{Snyder2016}. For an external Cox model, suppose the corresponding continuous hazard function for patient $i$ with covariate $\textbf{X}_i$ at time $t$ is given by $\tilde{\lambda}(t|\textbf{X}_i)=\tilde{\lambda}_0(t)\exp{(\textbf{X}^\top_{i} \tilde{\boldsymbol\beta})}$, where $\tilde{\boldsymbol\beta}$ is the regression parameters and $\tilde{\lambda}_0(t)$ is the baseline hazard function. When the continuous times are grouped into disjointed time intervals, as defined on the internal data,  $(0, c_1), [c_1, c_2), \dots,$ and $ [c_{K-1}, c_K)$, the predicted discrete hazard for subject $i$ in the $k$-th interval is given by
\begin{equation*}
\tilde{P}(T_i\in [c_{k-1}, c_k)|T_i\geq c_{k-1}, \textbf{X}_i) = 1-(1-\tilde{\lambda}_k)^{\exp{(\textbf{X}^\top_{i} \tilde{\boldsymbol\beta})}}, \text{ } k=1, \cdots, K, \text{ and } i=1, \cdots, n,
\label{eq:external_cox_prob}
\end{equation*}
where $\tilde{\lambda}_k=1-\exp{\{-\int_{c_{k-1}}^{c_k}\tilde{\lambda}_0(u) du\}}$ is the discrete baseline hazard.

\newpage
\textit{Example 2 (Predicted hazards based on discrete survival models).} 
Discrete failure time models are the natural choice when developing prediction models with intrinsically discrete measurements or grouped data. 
Suppose an external discrete failure time model with parameters $(\tilde{\bm{\gamma}},\tilde{\bm{\beta}})$ is available to be integrated with the internal data. The corresponding predicted hazard for subject $i$ in the internal data at time $t_k$ is
\begin{equation*}
\tilde{\lambda}(t_k;\textbf{X}_{i}) = g(\tilde{\gamma}_k + \textbf{X}^\top_{i} \tilde{\boldsymbol\beta}), \text{ } k=1, \cdots, K, \text{ and } i=1, \cdots, n.
\label{eq:external_disc_prob}
\end{equation*}


\section{Kullback-Leibler-Based Discrete Failure Time Models}
\label{s:method}

Given both an external survival prediction model and internal time-to-event data, in constructing the integrated prediction model, we have two possibly conflicting goals: (1) fit the internal data, where we maximize the log-likelihood of the internal discrete failure time model; and (2) fit the external model, where we propose a discrete hazard-based KL discrimination information that measures the discrepancy between the external model and the internal data (Section~\ref{s:new_KL}). 
We then develop a KL-based integration framework (Section~\ref{s:inte}), which incorporates external information to the internal discrete failure time model and balances the fit to the external model and the internal data.


\subsection{Discrete hazard KL discriminatory information}
\label{s:new_KL}

A key ingredient of the proposed procedure is the discrete hazard-based KL information. Although the classical density-based KL information has been applied to data integration problems for binary classification \citep{Liu2003,Schapire2005, plasso}, one major challenge for time-to-event data is that a censored outcome only provides partial information that the unobserved failure time is greater than the observed censoring time. This precludes the application of the classical density-based KL information. 

To incorporate both failed and censored information, we propose a discrete hazard-based KL information. As shown in Section~\ref{s:notation}, first, the discrete hazard is a conditional probability; second, under discrete failure time models, both the survival function \eqref{eq:survival} and the density function \eqref{eq:density}, and hence, the likelihood function \eqref{eq:likelihood} can be fully represented by the hazard functions. These motivate us to define KL information based on the discrete hazards. 

Suppose $\tilde{\lambda}$ and $\lambda$ are hazard functions corresponding to the external failure time distribution $\tilde{P}$ and the internal distribution $P_{\boldsymbol{\theta}}$ (with parameter $\boldsymbol{\theta}$), respectively. 
Here $\tilde{P}$ is considered to be given, and only $P_{\boldsymbol{\theta}}$ is indexed by $\boldsymbol{\theta}$.
Conditional on that patient $i$ is at risk at time $t_k$, we define the discrete hazard KL discrimination information between $\tilde{P}$ and $P_{\boldsymbol{\theta}}$ at time $t_k$ as
\begin{eqnarray}
 d_{KL}(\tilde{P} \parallel P_{\boldsymbol{\theta}};\textbf{Z}_{i}, t_k) = {}\tilde{\lambda}(t_k; \textbf{X}_{i}) \log\left\{\frac{\tilde{\lambda}(t_k; \textbf{X}_{i})}{\lambda(t_k; \textbf{Z}_{i})}\right\} +{} \{1-\tilde{\lambda}(t_k; \textbf{X}_{i})\} \log\left\{\frac{1-\tilde{\lambda}(t_k; \textbf{X}_{i})}{1-\lambda(t_k; \textbf{Z}_{i})}\right\}. 
 \label{eq:d}
\end{eqnarray}
Note that $d(\tilde{P} \parallel P_{\boldsymbol{\theta}};\textbf{Z}_{i}, t_k)$ can also be considered as the KL information for longitudinal binary outcomes. That is, the event status at each time point for which the patient is at risk  can  be  viewed  as  a binary outcome, $\Delta_{ik}\overset{d}{=} I(T_i=t_k|T_i\geq t_k, \textbf{Z}_{i})$, where $d$ stands for distribution.  
The accumulated KL information in the sequence of longitudinal binary outcomes over subjects in the internal data is given by 
\begin{equation}
D_{KL}(\tilde{P} \parallel P_{\boldsymbol{\theta}})=\sum_{i=1}^n \sum_{k=1}^K Y_i(t_k)d_{KL}(\tilde{P} \parallel P_{\boldsymbol{\theta}}; \textbf{Z}_{i}, t_k).
\end{equation}

The proposed discrete hazard KL discriminatory information provides a natural metric for measuring the discrepancy between the external model and the internal data, incorporating both failed and censored observations. We will show in Section~\ref{s:estimation} that both the proposed discrete hazard-based KL information and the resulting objective function of the KL-based integration procedure retain similar forms as the classical discrete failure time models, and hence, the corresponding estimation is computationally efficient and can be easily extended to high-dimensional problems.


\subsection{KL-based integration framework}
\label{s:inte}
 To balance the trade-off between the external model and the internal data, we maximize the log-likelihood of the internal discrete failure time model, while at the same time keeping the accumulative KL discrimination information between the internal data and the external model small. This can be achieved by maximizing a penalized log-likelihood 
\begin{equation}
  \ell_{\eta}(\bm{\theta})
= \ell(\bm{\theta})- \eta  D_{KL}(\tilde{P}\parallel P_{\bm{\theta}}),
\label{eq:penalized-loglik0}  
\end{equation}
where $\eta$ is a tuning parameter weighing the relative importance of the external model to the internal data. Thus, the KL-based integration controls the relative weight of the external information, assigning larger weights to more compatible ones and diminishing less relevant ones. In the extreme case of $\eta = 0$, the penalized log-likelihood reduces to the log-likelihood based on the internal data. When $\eta = \infty$, maximizing the penalized log-likelihood is equivalent to the optimization problem of minimizing $D_{KL}(\tilde{P}\parallel P_{\bm{\theta}})$. 

The optimal tuning parameter $\eta$ can be determined by cross-validation to minimize the predictive deviance \citep{model_select, Tutz16}, i.e. the negative log-likelihood of the fitted model evaluated on the validation data, which measures how far apart the working model is 
from the underlying model. 
The proposed method can also be easily extended to incorporate multiple external models, identifying the most compatible external data sources and diminishing less relevant ones; further details are provided in the Supporting Information.

\subsection{Optimality}
\label{s:optimalizty}
Here we give the optimality property of the proposed KL-based integration procedure. It can be shown that, given $\eta$, the $\ell_{\eta}(\bm{\theta})$ provides an optimal estimate that minimizes the convex sum of the KL information between the working model and two extremes: one in which no external information is used and the other in which no internal data is used. We present this optimality property in Proposition 1.

\textbf{Proposition 1}
{\it  Let $P_0$ be the probability distribution corresponding to the saturated model of longitudinal binary outcomes in the internal data. For a given $\eta$, the model $P_{\bm{\theta}}$ that minimizes 
 \begin{align}
&(1-\eta_0) D_{KL}(P_0\parallel P_{\bm{\theta}}) + \eta_0  D_{KL}(\tilde{P}\parallel P_{\bm{\theta}})
\label{eq:penalized-kl}  
\end{align}
is the KL-integrated model in (\ref{eq:penalized-loglik0}), where $\eta_0 = \eta/(1+\eta)$. 
}

A proof is provided in the Supporting Information. Proposition 1 shows that the proposed KL-based integration procedure provides an optimal combination of the external model and the internal data, which justifies the form of the proposed KL-based integration objective function.


\subsection{Estimation}
\label{s:estimation}

Proposition 2 below shows that $ \ell_{\eta}(\boldsymbol{\theta})$ has a similar form as the log-likelihood of discrete failure time models using the internal data only. 
 Thus, the parameter estimation can be easily obtained by classical procedures such as the Newton-Raphson procedure \citep{Kalbfleisch2002} for low-dimensional settings. An extension to high-dimensional settings is provided in the next subsection.
 
 \noindent
\textbf{Proposition 2}
{\it
The accumulated KL information is given by
\begin{equation*}
  D_{KL}(\tilde{P}\parallel P_{\bm{\theta}}) = - \sum_{i=1}^n\sum_{k=1}^K Y_i(t_k)\left[\tilde\lambda(t_k;\textbf{X}_{i}) \log\left\{\frac{g(\gamma_{k}  + \textbf{Z}^\top_{i} \boldsymbol\beta)}{1-g(\gamma_{k}  + \textbf{Z}^\top_{i} \boldsymbol\beta)}\right\}  + \log\{1-g(\gamma_{k}  + \textbf{Z}^\top_{i} \boldsymbol\beta)\}\right]+\Psi,
\end{equation*}
where $\Psi$ is a constant not involving $\bm{\theta}$.
Therefore, $\ell_{\eta}(\bm{\theta})$ in \eqref{eq:penalized-loglik0} is proportional to the following objective function
\begin{equation}
 \sum_{i=1}^n\sum_{k=1}^{K} Y_i(t_k)\left[ 
\frac{\delta_i(t_k)+\eta\tilde\lambda(t_k;\textbf{X}_{i})}{1+\eta}
\log\left\{\frac{g(\gamma_{k}  + \textbf{Z}^\top_{i} \boldsymbol\beta)}{1-g(\gamma_{k}  + \textbf{Z}^\top_{i} \boldsymbol\beta)}\right\} + \log\{1-g(\gamma_{k}  + \textbf{Z}^\top_{i} \boldsymbol\beta)\} \right],
\label{eq:penalized-loglik}   
\end{equation}
which retains a form similar to the  log-likelihood of discrete failure time models using the internal data only.
}

A proof of Proposition 2 is provided in the Supporting Information. In practice, the proposed integration procedure can be implemented with various link functions. Common choices for the link function include complementary log-log link, $h(u)=log\{-log(1-u)\}$, which leads to the grouped relative risk model \citep{kalbfleisch1973marginal}; log link,  $h(u)=log(u)$, which is the discrete relative risk model \citep{prentice2003mixed}; and logit link, $h(u)=log\{u/(1-u)\}$, which is the discrete logistic model \citep{cox1972regression}. 

In most applications, the discrete hazard at each time point is relatively small, and all three link functions provide similar results. Nevertheless, the discrete relative risk model retains the relative risk interpretation of the multiplicative factor  and can be applied for recurrent events, while the discrete logistic model is a natural choice for longitudinal binary outcomes. 
In addition, as shown in Example 1 of Section~\ref{sub:external_model}, the grouped relative risk model can be obtained by grouping the continuous time, and hence, is uniquely appropriate for integrating published Cox models. 

\subsection{An Extension: KL-based Coordinate-wise Gradient Boosting}
\label{s:extend}

To achieve variable selection for high-dimensional problems, we extend the coordinate-wise gradient boosting  \citep{Buhlmann2003Boosting, Buhlmann2007Boosting, Buhlmann:2011} and propose an iterative KL-based variable selection procedure. Specifically, in  the  outer layer  of  each  iteration, given the current estimate of $\bm{\beta}$, we update the transformed baseline hazard $\bm{\gamma}$ by a one-step Newton procedure. In the inner layer, given the current estimate of $\bm{\gamma}$, to achieve variable selection on $\bm{\beta}$, we consider a first-order Taylor’s expansion: 
\begin{eqnarray}
 \ell_{\eta}(\boldsymbol{\widehat\beta}+\alpha \boldsymbol \mu) = \ell_{\eta}(\boldsymbol{\widehat\beta})+\alpha \triangledown \ell_{\eta}(\widehat{\boldsymbol\beta})^T\boldsymbol \mu+ \frac{1}{2}\alpha^2\boldsymbol \mu^T\triangledown^2 \ell_{\eta}(\widehat{\boldsymbol\beta}+w\boldsymbol \mu)\boldsymbol \mu,
 \nonumber
\end{eqnarray}
where $\boldsymbol{\widehat\beta}$ is the current estimate; $\boldsymbol \mu$ is the update direction of $\boldsymbol\beta$; $\alpha$ is a small positive value; $w\in[0,\alpha]$; $\triangledown \ell_{\eta}$ and $\triangledown^2 \ell_{\eta}$ are the corresponding gradient vector and Hessian matrix, respectively; and the term $\triangledown \ell_{\eta}(\widehat{\boldsymbol\beta})^T\boldsymbol \mu$  is the directional derivative along  $\boldsymbol \mu$.
If  $\triangledown \ell_{\eta}(\widehat{\boldsymbol\beta})^T\boldsymbol \mu >0$, the direction $\boldsymbol \mu$ is an ascent direction  of $\boldsymbol\beta$  to increase   $\ell_{\eta}(\boldsymbol\beta)$.
We  identify an update
direction (with a unit norm), along which
$\ell_{\eta}(\boldsymbol\beta)$
 ascends most rapidly.
This motivates a steepest ascent update direction, 
  \begin{eqnarray}
 \boldsymbol \mu^{\star}= \argmax_{\boldsymbol \mu} \{ \triangledown \ell_{\eta}(\widehat{\boldsymbol\beta})^T\boldsymbol \mu ~\big |~ ||\boldsymbol \mu||_1=1\},
 \nonumber
\end{eqnarray}
where $||\cdot||_1$ is the $\ell_1$ vector norm on $\mathbb{R}^{p}$.  
With the generalized Cauchy-Schwarz inequality,
$
\triangledown \ell_{\eta}(\widehat{\boldsymbol\beta})^T\boldsymbol \mu
 \leq  ||\triangledown \ell_{\eta}(\widehat{\boldsymbol\beta})||_{\infty} ||\boldsymbol \mu||_1,
 \nonumber
 $ where $||\cdot||_{\infty}$ is the $\ell_{\infty}$ vector norm. 
 Thus, the coordinate-wise steepest ascent direction maximizing the directional derivative is given by
 \begin{eqnarray}
\boldsymbol \mu^{\star}= \argmax_{\boldsymbol \mu} \{ \triangledown \ell_{\eta}(\widehat{\boldsymbol\beta})^T\boldsymbol \mu ~\big | ~ ||\boldsymbol \mu||_1=1\}=(0, \ldots, 0, \widehat{\boldsymbol\mu}_{j^{\star}}, 0, \ldots, 0)^T,
 \nonumber
\end{eqnarray}
 where
$
j^{\star}=\argmax_j \left(\big|\triangledown \ell_{\eta}(\widehat{\boldsymbol\beta})_j\big|\right)
$
and $
\widehat{\boldsymbol\mu}_{j^{\star}}=  \mbox{sign}(\triangledown \ell_{\eta}(\widehat{\boldsymbol\beta})_{j^{\star}}).
$
We summarize the KL-based component-wise gradient boosting algorithm as follows:
 \begin{enumerate}
 \vspace{-0.3cm}
 \item[(a)] Initialize $\boldsymbol {\widehat{\theta}}^{(0)}=\textbf{0}$.  For $s=1,2,3, \ldots$, iterate the following: 
 \item[(b)] Given  $\widehat{\boldsymbol\beta}^{(s-1)}$, update the baseline hazards $\bm{\gamma}$ by a one-step Newton procedure.
  \item[(c)] 
  Given $\widehat{\boldsymbol\gamma}^{(s)}$,
  update  $\bm{\beta}$  by
$
 \boldsymbol {\widehat  \beta}^{(s)}_{j^{\star}}= \boldsymbol {\widehat  \beta}^{(s-1)}_{j^{\star}}+\nu ~ \widehat{\boldsymbol\mu}_{j^{\star}},
$ where $\nu$ is a small positive value (e.g., 0.01) controlling the learning rate. 
\vspace{-0.25cm}
\end{enumerate}

The component-wise algorithm ranks the importance of the predictors and measures how fast the penalized log-likelihood will increase by including each predictor. At each iteration, only  one  variable  is  updated.   Thus,  variable  selection  is  automatically  achieved  for an  optimal  number  of  iterations,  determined  by early  stopping \citep{Zhang2005, Mayr2012}; i.e. the relative change
in the penalized log-likelihood is less than a  threshold (e.g. $10^{-3}$).

Following the strategy in \citet{He2021}, the proposed KL-based boosting procedure can be interpreted from the perspective of the Minorization-Maximization algorithm, which reaffirms that the estimates in each iteration serve as refinements of the previous step and helps clarify the numerical advantage of the proposed procedure.
Specifically,  given the current estimate of $\bm{\gamma}$, we consider a minority surrogate function,
\begin{align*}
\phi(\boldsymbol\beta|\widehat{\boldsymbol\beta}) =  \ell_{\eta}(\widehat{\boldsymbol\beta}) + \triangledown \ell_{\eta}(\widehat{\boldsymbol\beta})^T (\boldsymbol\beta-\widehat{\boldsymbol\beta})-\frac{1}{2\nu}(\boldsymbol\beta-\widehat{\boldsymbol\beta})^T  (\boldsymbol\beta-\widehat{\boldsymbol\beta}),
\end{align*}
where $\nu$ is a small positive value to be specified.  
With $\phi(\boldsymbol{\widehat\beta}|\boldsymbol{\widehat\beta})=\ell_{\eta}(\boldsymbol{\widehat\beta})$, Proposition 3 below shows that, given a suitable  $\nu$,  $\phi(\boldsymbol\beta|\widehat{\boldsymbol\beta}) \leq  \ell_{\eta}(\boldsymbol\beta)$ for all $\boldsymbol\beta$. Thus, $\phi(\boldsymbol\beta|\widehat{\boldsymbol\beta})$ serves as a minority surrogate function of $\ell_{\eta}(\boldsymbol\beta)$, 
which is separable across covariate coefficients and, hence, reduces a high-dimensional optimization problem to simpler ones. The coordinate-wise gradient update maximizes $\phi(\boldsymbol{\beta}|\boldsymbol{\widehat\beta})$ subject to the constraint that only one variable is updated at each iteration. The resulting estimates of $\boldsymbol{\beta}$ ensure the ascent property and thus serve as refinements of the previous step. The detailed proofs are provided in the Supporting Information.
 
\textbf{Proposition 3 (Ascent Property)}

{\it
For $\nu>0$ satisfying
\begin{align}
\sup_{ \{\boldsymbol\beta: \ell_{\eta}(\boldsymbol\beta) \geq \ell_{\eta}(\boldsymbol\beta^{(0)})\}} \left( \lambda_{max}\left\{- \triangledown^2 \ell_{\eta}(\boldsymbol\beta)\right\} \right) < 1/\nu,
\label{eq:rate}
\end{align}
then  $\phi(\boldsymbol\beta|\widehat{\boldsymbol\beta}) \leq  \ell_{\eta}(\boldsymbol\beta)$ for all $\boldsymbol\beta$, where $\lambda_{max}(\cdot)$ represent the largest eigenvalues. 
   }

\section{Simulation}
\label{s:sim}

To assess the performance of the proposed KL-based integration framework for discrete survival problems, we conducted simulations for both low-dimensional and high-dimensional settings. Generally, the proposed modeling procedure works for discrete time-to-event models with logit, complementary log-log and log link functions. Here, we focus on the logit link function (i.e. discrete logistic regression) as an illustrating example. The tuning parameter $\eta$ of the proposed KL-based integration method was selected by 5-fold cross validation, with the average predictive deviance  as the metric of the model performance. We then evaluated the model performance on a testing dataset, which was independently simulated from the same distribution as the internal data. The simulation studies were replicated 100 times.

\subsection{KL-based discrete logistic models for low-dimensional settings}
\label{sub:low_dim}

In the low-dimensional settings, the internal data were generated with $10$ predictors, $\bm{Z}_l=\{\bm{Z}_{1}, \dots,\bm{Z}_{10} \}$, where continuous covariates $\{\bm{Z}_{1}, \dots,\bm{Z}_{5} \}$ came  from  a  multivariate  normal  distribution with zero mean, unit variance and  a first-order autoregressive (AR1) correlation structure with the  auto-correlation  parameter 0.5; and $\{\bm{Z}_{6}, \dots,\bm{Z}_{10} \}$ were binary variables with $Pr(\bm{Z}_i=1)=0.5$. Failure times were generated from a discrete logistic model with parameters $(\bm{\gamma},\bm{\beta})$. In particular, four simulation settings regarding censoring rate and sample size of the internal cohort were explored. We first considered a high censoring rate setting with covariate effects as $\bm{\beta}=(-2,1,-2,-3,1,-4,1,-3,-4,1)^\top$.
\begin{itemize}
    \item[(I).] Censoring rate $90\%$; Sample size: 300.
\end{itemize}
Then we considered three low censoring rate settings with covariate effects as \\
$\bm{\beta}=(2,-1,2,3,-1,4,-1,3,4,-1)^\top$, where we demonstrated the effects of the censoring rate and sample size of the internal data on the performance of the proposed method.
\begin{itemize}
    \item[(II).] Censoring rate $20\%$; Sample size: 75.
    \item[(III).] Censoring rate $20\%$; Sample size: 100. 
    \item[(IV).] Censoring rate $40\%$; Sample size: 100.    
\end{itemize}
For all four simulation settings, the logit-transformed baseline hazards were set as $\bm{\gamma}=(-6,-6,-6, -4.5, -4.5, -4.5, -3.0, -3.0, -1.5, -1.5)^\top$. Censoring times were generated from a discrete uniform distribution with varying upper bounds to control for different censoring proportions. The external model was estimated by a discrete logistic regression model on a large external cohort with sample size 10,000. 
The underlying distribution of the external cohort was the same as the internal data, while the observed external samples may only contain a subset of covariates. Specifically, 
we considered the following four scenarios for the external models, including a setting with multiple external models:

\begin{itemize}
    \item[(a).] The external cohort contained the full set of covariates:   $\bm{X}=\{\bm{Z}_{1}, \dots,\bm{Z}_{10}\}$.
   \item[(b).] The external cohort only included a subset of covariates:  $\bm{X}=\{\bm{Z}_{1}, \dots, \bm{Z}_{8}\}$.
   \item[(c).] The external cohort only included a smaller subset of covariates:  $\bm{X}=\{\bm{Z}_{1}, \bm{Z}_{2}, \bm{Z}_{6}, \bm{Z}_{7}\}$.
   \item[(d).] Two external models were available, with external cohort 1 including eight covariates:  $\bm{X}_{1}=\{\bm{Z}_{1}, \dots, \bm{Z}_{8}\}$, and external cohort 2 including four covariates: $\bm{X}_{2}=\{\bm{Z}_{1}, \bm{Z}_{2}, \bm{Z}_{6}, \bm{Z}_{7}\}$.   
\end{itemize}

In addition to the model fitted by the proposed KL-based discrete failure time model, competing methods considered included: (1) the external model by applying parameters of the external model to the internal data directly,
(2) the internal model fitted on the internal data only, 
and (3) the stacked model fitted by extending the stacked regression model introduced in \citet{debray14} to discrete time-to-event problems, which treats predictions of each external model as a predictor in the integrated model. 

\begin{figure}[htbp]
\begin{center}
\centerline{\includegraphics[width=14cm]{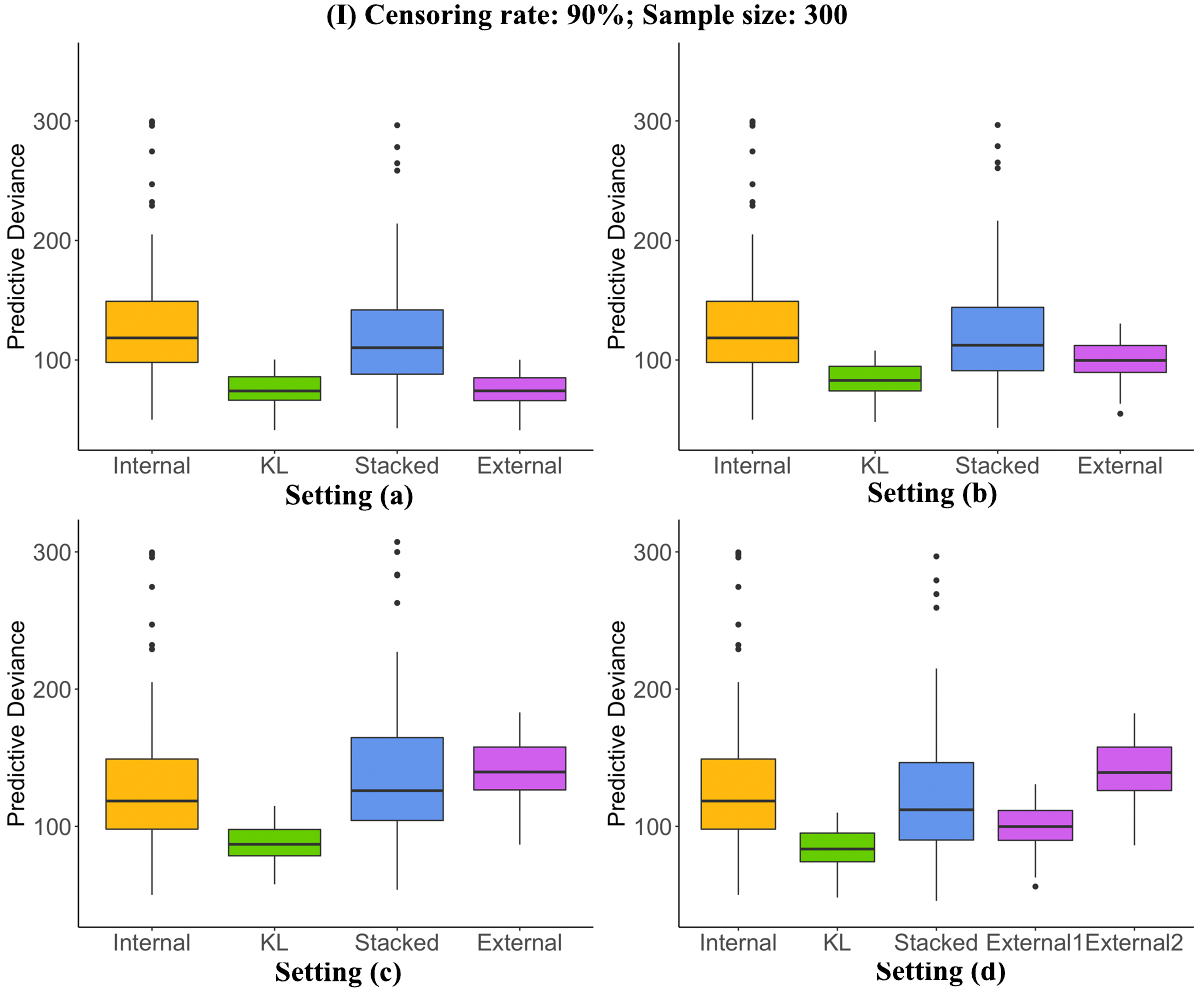}}
\end{center}
\caption{Simulation results of internal data setting (I) across external model settings (a)-(d). Lower predictive deviance indicates better prediction performance.
\label{fig:kl_simu1}}
\end{figure}

Figure \ref{fig:kl_simu1} shows the simulation results of the internal setting (I) across external models (a)-(d). Compared with the internal, external and stacked models across different external scenarios, the proposed KL-based integration procedure achieved the best performance. Specifically, under the situations where the external models were mis-specified and only included a subset of internal variables, the proposed method made the best prediction with the smallest predictive deviance. When the external model was the underlying true model and contained all the variables available in the internal data, the KL-based integration method achieved comparable performance to the external model. Moreover, the interquantile range (IQR) of the KL model was narrower than those of the internal and stacked models, which indicates that incorporating information from external models by the proposed KL-based integration procedure improves the prediction stability.

\begin{figure}[htbp]
\begin{center}
\centerline{\includegraphics[width=19cm]{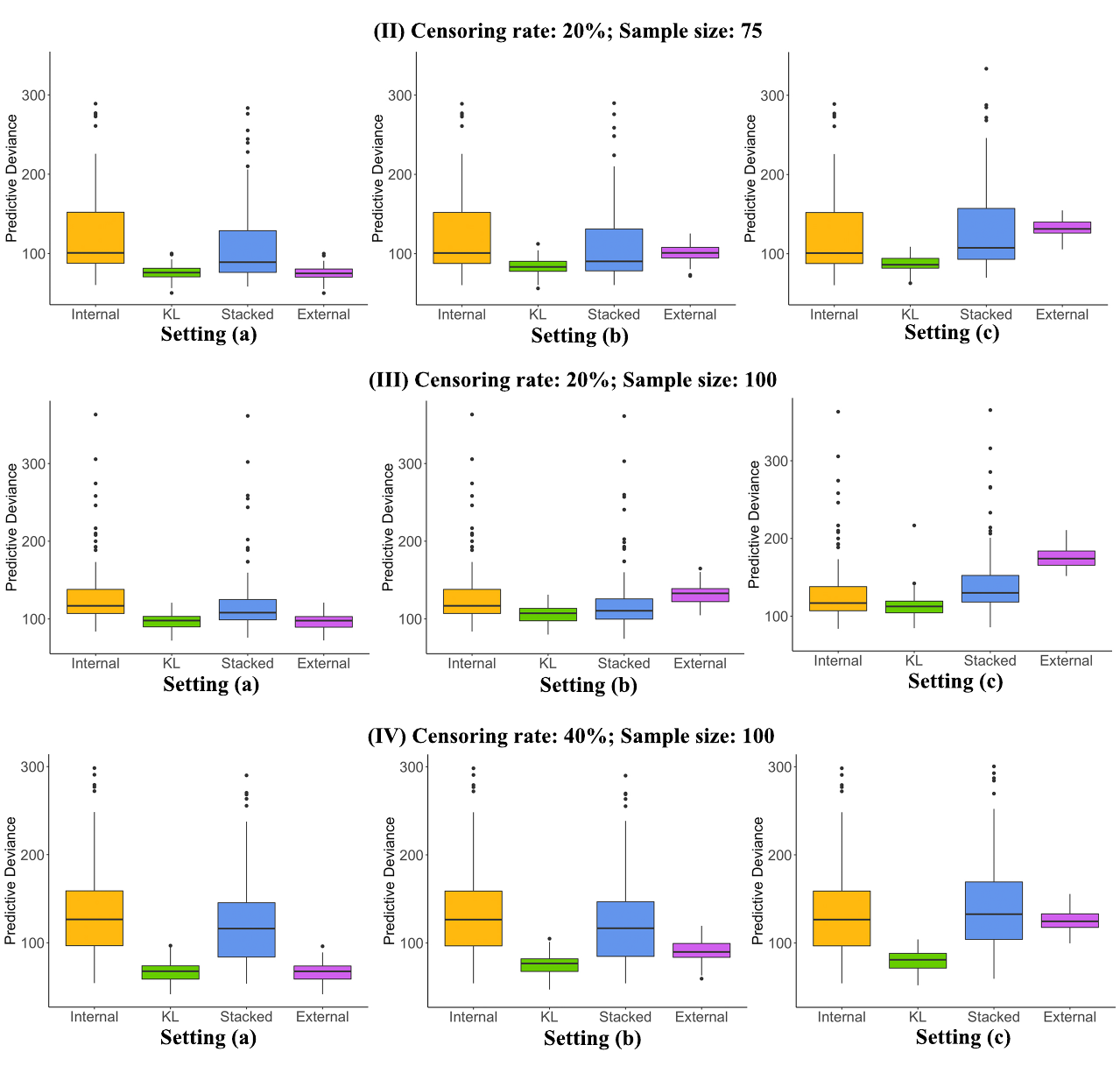}}
\end{center}
\caption{Simulation results of internal data settings (II)-(IV) across external model settings (a)-(c). Rows 1-3 correspond to internal settings (II)-(IV). Columns 1-3 correspond to external model settings (a)-(c). Lower predictive deviance indicates better prediction performance.
\label{fig:kl_simu2}}
\end{figure}

As shown in Figure \ref{fig:kl_simu2}, the predictive deviance of our method was lower and more stable compared to other methods under various internal settings regarding censoring rate and sample size. As the sample size decreased or the censoring rate increased, the prediction performance of the internal and stacked models became less accurate and less stable, while the performance of the proposed method retained its superiority. The proposed method tended to select a larger $\eta$ when the internal data were more relevant to that of the external model. In addition, the proposed KL-based integration method diminished misleading or irrelevant external information by selecting an extremely small $\eta$ or setting it to be 0 (Figure S1).

\subsection{KL-based gradient boosting for high-dimensional settings}

In high-dimensional settings, two sets of simulations were conducted to assess the performance of the proposed method on the improvement of the variable selection and the prevention of overfitting. 

In the first set of simulations, the internal data were generated with 200 subjects and 400 predictors, where the predictors followed a multivariate normal distribution with a block-diagonal covariance structure ($200$ independent blocks, each with 2 predictors). Within each block the variables  followed a bi-variate normal distribution with the correlation  parameter 0.75. We generated outcomes such that 5 predictors were associated with the failure times. Without loss of generality, we assumed $\{\bm{Z}_{1}, \bm{Z}_{3}, \bm{Z}_{5}, \bm{Z}_{7}, \bm{Z}_{9}\}$ from the first 5 blocks were true signals, with covariate effects as $(0.08, 0.08, -0.08, -0.08, 0.08)^\top$. Hence, $\{\bm{Z}_{2}, \bm{Z}_{4}, \bm{Z}_{6}, \bm{Z}_{8}, \bm{Z}_{10}\}$ were noise signals correlated with the truth.
The logit-transformed baseline hazards were set as $(-0.01, -0.02, -0.03, -0.04, -0.05)^\top$. The external model was estimated from a large external cohort with sample size 10,000. We considered the following two scenarios for the external model:
\begin{itemize}
    \item[(e).] The external study identified not only a subset of true signals, but also two noise signals: $\bm{X}=\{\bm{Z}_{1}, \bm{Z}_{3}, \bm{Z}_{5}, \bm{Z}_{8}, \bm{Z}_{11}\}$, where $\bm{Z}_{8}$ was correlated with true signal $\bm{Z}_{7}$, and $\bm{Z}_{11}$ was independent of true signals.
    \item[(f).] The external study identified the full set of true signals: $\bm{X}=\{\bm{Z}_{1}, \bm{Z}_{3}, \bm{Z}_{5}, \bm{Z}_{7}, \bm{Z}_{9}\}$.
\end{itemize}
Scenario (f) mimicked the situation where the external model provided relevant information to the internal study, while the external model in scenario (e) contained a mixture of relevant and irrelevant information. The empirical selection proportion path, which plots the empirical selection proportion of each variable of 100 simulation replicates along with iteration steps, was used as the performance metric of variable selection in the high-dimensional setting. 

\begin{figure}[htbp]
\begin{center}
\centerline{\includegraphics[width=12cm]{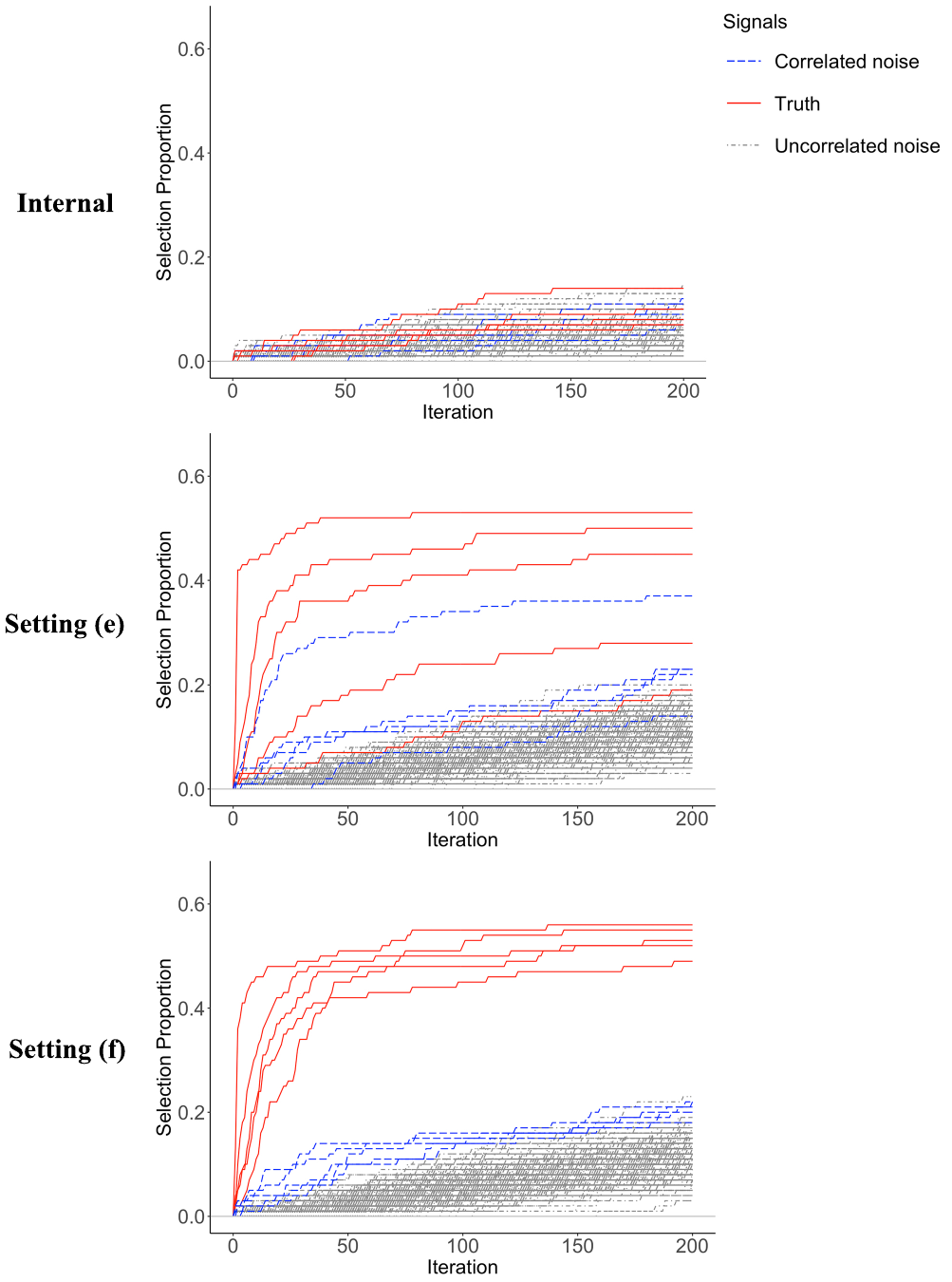}}
\end{center}
\caption{Comparison of performance of variable selection under different simulation settings. Internal: classical gradient boosting model fitted by internal data only; Setting (e): KL-based gradient boosting model integrating external model (e) with internal data; Setting (f): KL-based gradient boosting model integrating external model (f) with internal data. External model (e) contains both relevant and irrelevant information, while external model (f) only contains relevant information to the internal study. The empirical selection proportion path plots the selection proportion of each variable under 100 simulation replicates along with iteration steps. Red solid lines: paths of true signals; Blue dashed lines: paths of noise signals correlated with truth; Gray dashed lines: paths of noise signals uncorrelated with truth. 
\label{fig:kl_high_dim}}
\end{figure}

We compared the proposed KL-based gradient boosting model with the internal model fitted by the gradient boosting \citep{Buhlmann2007Boosting} using the internal data only. When the signal to noise ratio is low, it is particularly difficult to tease apart the true signals from the noises using the internal data only. In contrast, as shown in Figure \ref{fig:kl_high_dim}, the KL-based gradient boosting improved the variable selection performance by incorporating external information. Specifically, when the external study identified the full set of true signals, all the informative variables (with non-zero coefficient) were distinguished with the proposed KL method. Even when the external study only identified partially true signals, informative variables were more likely to be selected than they were with the internal data only. The proposed method favored more relevant external information with a better selection proportion path.

\begin{figure}[htbp]
  \centerline{\includegraphics[width=18cm]{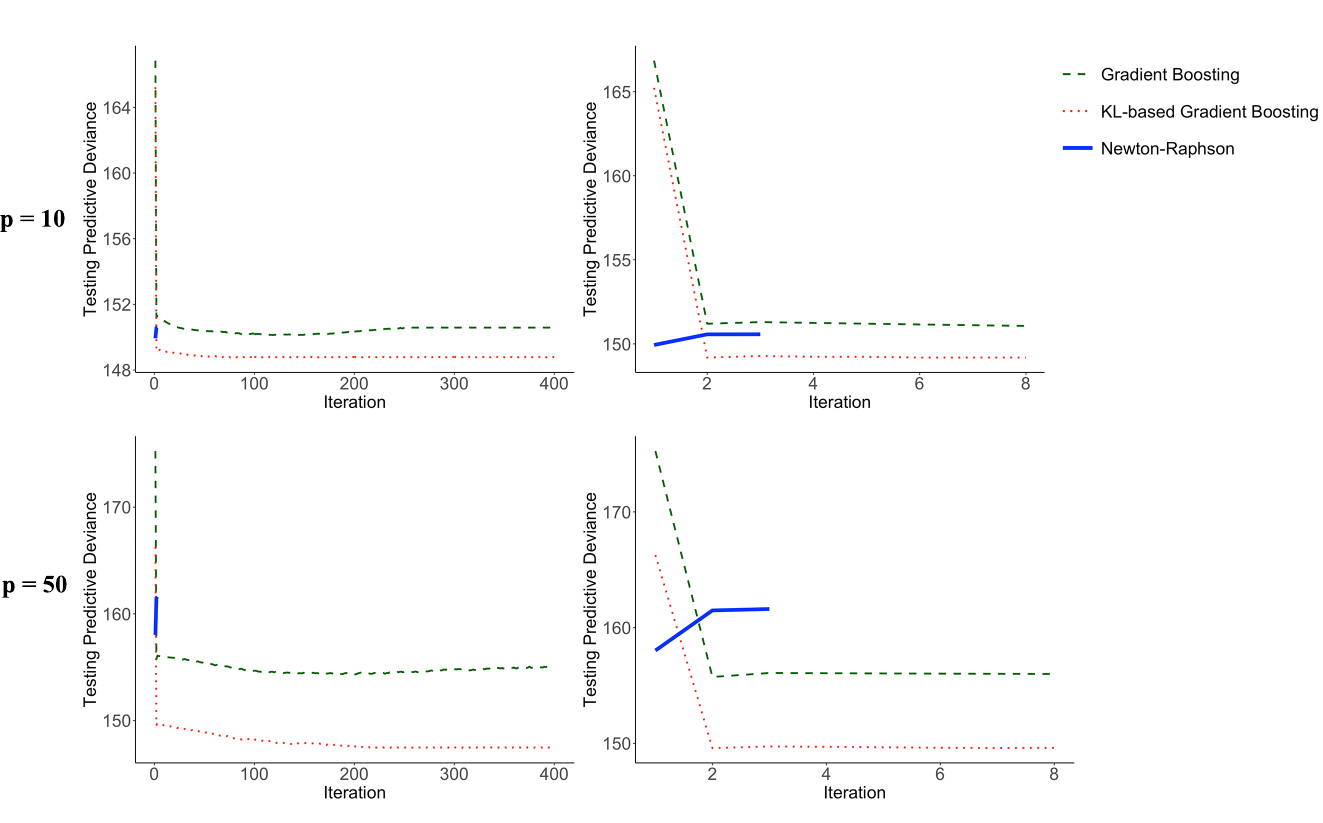}}  
  \caption{Progression of testing predictive deviance along with iteration steps under different numbers of predictors of the internal data ($p=10$ and $p=50$). Green dashed lines: Gradient boosting with internal data only; Red dotted line: KL-based gradient boosting integrating external model with internal data;  Blue solid lines: Newton-Raphson with internal data only. The figures on the right panel were plotted by zooming the first 8 iterations of the figures on the left.}
\label{fig:fig4}
\end{figure}

In the second set of simulations, we simulated the internal data following the same generating mechanism in the first set of simulations, but with the number of predictors varying from 10 to 50. The true signals were set as $\{\bm{Z}_{1}, \bm{Z}_{3}, \bm{Z}_{5}, \bm{Z}_{7}, \bm{Z}_{9}\}$, and we assumed that the external study identified a mixture of truths and noises ($\bm{X}=\{\bm{Z}_{1}, \bm{Z}_{3}, \bm{Z}_{5}, \bm{Z}_{6}, \bm{Z}_{8}\}$). In addition to the proposed KL-based gradient boosting, we also explored the performances of Newton-Raphson and classical gradient boosting. Figure \ref{fig:fig4} shows the progression of testing predictive deviance along with iterations of the aforementioned three algorithms under different numbers of predictors. The proposed KL-based gradient boosting outperformed the other two algorithms in all simulation settings. In particular, the proposed method attained the lowest testing predictive deviance after a few iterations. Moreover, our method did not suffer from overfitting issues; however, both Newton-Raphson and the classical gradient boosting algorithms tended to overfit the data after several iterations. 

\section{Real Data Analysis}
\label{s:real}
We then applied the proposed method to integrate a published SRTR prediction model with a small-sized internal dataset to improve the post-transplant outcome prediction for patients with ESRD. 

The SRTR post-transplant risk adjustment model was developed by the SRTR contractor using the data from the U.S. Organ Procurement and Transplantation Network (OPTN). The SRTR model is derived from a Cox proportional hazards model used to predict 1-year kidney graft survival for adult recipients who received kidneys from deceased donors \citep{Snyder2016}. A total of 59 covariates were included in the SRTR model. In addition to model coefficients, baseline cumulative hazards are also reported, which enables the calculation of weekly predicted hazards. Although the SRTR model is derived from a massive kidney transplant cohort, it does not capture detailed clinical information, such as medical histories and complications not collected in the OPTN \citep{kasiske2019}. 

To investigate the effect of predictors outside the SRTR model on the 1-year post-transplant graft survival, an internal cohort from the University of Michigan Medical Center with $498$ ESRD patients who underwent kidney transplantation between January $2011$ and December $2015$ was analyzed. Failure time was defined as the time from transplantation to graft failure or death, whichever occurred first. Patient follow-up was censored 1-year post-transplant, and the overall censoring rate was $91.6\%$. Additional risk factors outside the SRTR model ($p=15$) in this study are shown in Table S1. Both the KL-based discrete survival model and KL-based gradient boosting model were applied in the analysis. Specifically, two scenarios using different sets of the internal variables were considered:

\begin{itemize}
    \item[(I):] All risk factors outside the SRTR model ($p=15$) were used in the analysis.
    \item[(II):] Only risk factors selected by the variable selection algorithms (indicated by $\ast$ in Table S1) were used in the analysis. 
\end{itemize}

\begin{figure}[htbp]
  \centerline{\includegraphics[width=18cm]{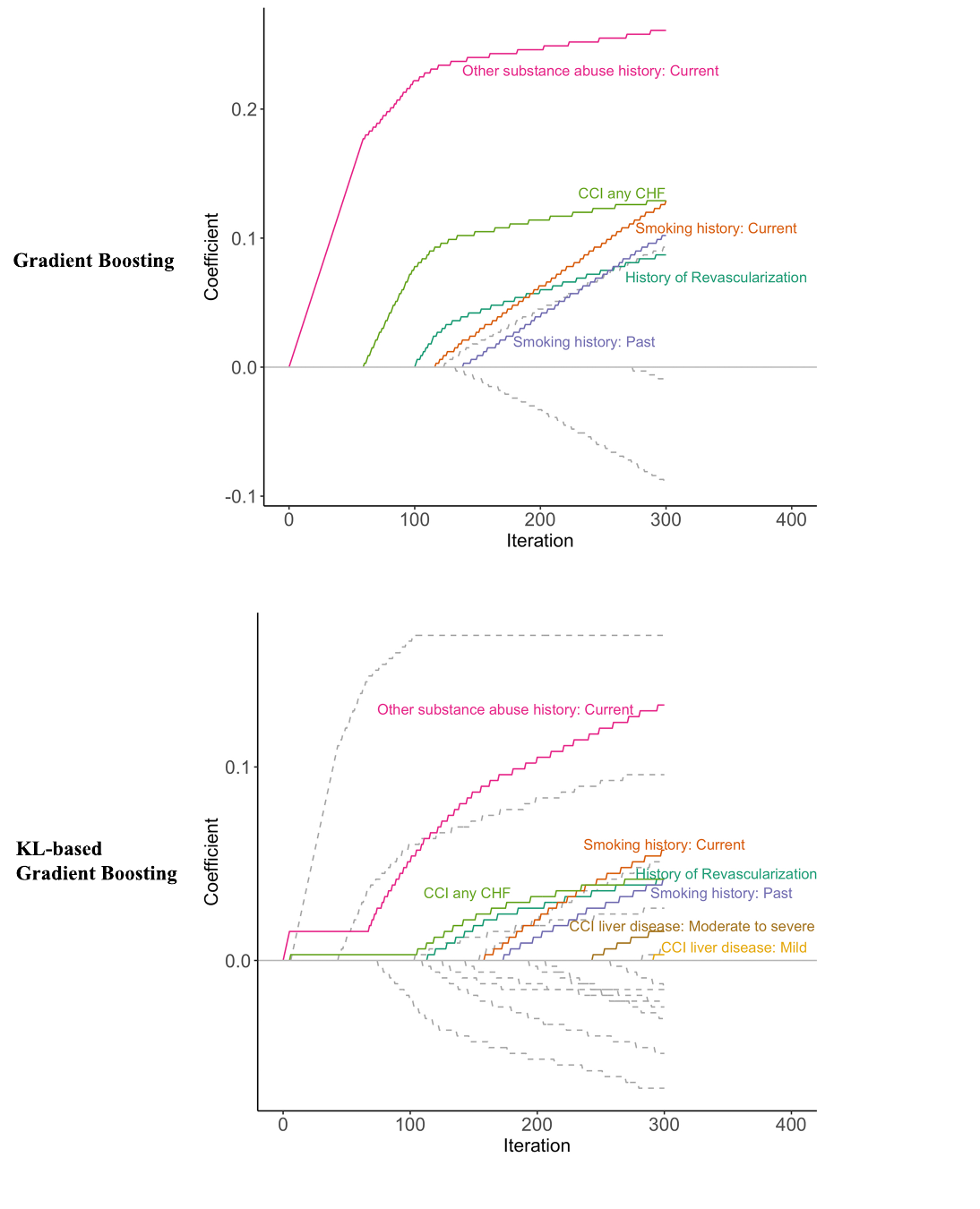}}  
  \caption{The solution paths for the kidney transplant study. Each colored line represents a coefficient path for a new risk factor outside the SRTR model. Each gray line represents a coefficient path for a risk factor included in the SRTR model. CCI = Charlson Comorbidity index; CHF = congestive heart failure.}
\label{fig:fig5}
\end{figure}

To incorporate the external SRTR model with the internal data, the complementary log-log link was applied in the analysis (as shown in Example 1 of Section~\ref{sub:external_model}). We randomly split the internal data into training and testing sets with a ratio of 4:1, developed prediction models on the training set, and evaluated the model performance on the leave-out testing set. A total of 10 independent random splits were implemented. 

As shown in Table \ref{tab:rl_rslt}, the proposed KL-based discrete failure time model achieved the best prediction performance under scenario I. Compared to the internal and stacked models, incorporating the SRTR model by KL-based method improved the accuracy and stability of prediction performance with smaller predictive deviance and standard error (SE). Models in scenario II performed better than those in scenario I, which indicates that variable selection also helps to improve the prediction performance. Thus, incorporating external information can help improve the prediction performance of the internal study, and the KL-based integration provides a more accurate and stable integrated model for prediction.

\begin{table}[htbp]
\begin{center}
\caption{Comparison of performance in predicting post-transplant graft survival in the kidney transplant study. Sample size of the internal data was $n=498$. Censoring rate of the internal data was $91.6\%$. Scenario I utilized all risk factors outside SRTR model ($p$=15). Scenario II only utilized risk factors selected by the corresponding variable selection algorithm ($p$=5 for KL-based gradient boosting and $p$=4 for gradient boosting). We randomly split the internal data into training and testing sets with a ratio of 4:1. The predictive deviance was evaluated on the leave-out testing set. A total of 10 independent random splits for the internal data were implemented. Smaller predictive deviance indicates better prediction performance. SE = standard error.  
\label{tab:rl_rslt}}
\vspace*{0.5cm}
\begin{tabular}{llcc}
\hline
{} &{} & {\bfseries Predictive} &{}\\
{\bfseries Scenario} &{\bfseries Method} & {\bfseries Deviance} &{\bfseries SE}\\
\hline
 \multirow{3}{*}{I}&KL-based discrete survival model & 13.56 & 5.59\\
  &Discrete survival model & 44.40 & 21.12\\
 &Stacked model  & 29.32 & 14.53\\
\hline 
 \multirow{2}{*}{II}&KL-based gradient boosting model & 13.53 & 5.55 \\
 &Gradient boosting model & 30.88 & 13\\
\hline 
\end{tabular}
\end{center}
\end{table}

Figure \ref{fig:fig5} shows the solution paths of the proposed KL-based gradient boosting and the classical gradient boosting on the internal kidney transplant data. In addition to smoking history, other substance abuse history, history of revascularization, and any congestive heart failure, the proposed KL-based gradient boosting selected liver disease as an important variable in terms of new risk factors outside the SRTR model. Liver diseases have been found as a risk factor associated with adverse outcomes after kidney transplantation \citep{Paramesh2012}. Moreover, the proposed KL-based gradient boosting algorithm also selected several predictors included in the SRTR model, while the gradient boosting algorithm only selected three. Since the SRTR model was built based on the large-scale OPTN dataset, 
incorporating the SRTR model by the proposed KL-based gradient boosting algorithm prevented model overfitting when the internal data sample size was small or the censoring rate of the internal data was high.  

\section{Discussion}
\label{s:discuss}

In this paper, we proposed a discrete hazard KL-based integration framework, which can incorporate published survival models from external data sources and improve the prediction performance of the internal time-to-event data. The proposed discrete hazard-based KL discrimination information  extends the classical KL discrimination information to discrete failure time data and provides a metric to
link the summary-level external information obtained from the published prediction models and the newly collected internal data. 
The proposed integration framework is robust to model mis-specification, automatically determining the relative similarity between external and internal data sources and incorporating more information from the more compatible external sources. Furthermore, the proposed method is flexible to  incorporate various  types  of  external  prediction  models,  including  machine learning  algorithms, as  long  as a sequence of predicted hazard functions or predicted survival probabilities across times can be provided by the external models. Both simulation and real data analyses demonstrate that the proposed method achieves improved prediction performance over existing methods.

While our work is motivated by kidney transplant, this method can also be applied to enhance prediction performance on other diseases, including rare ones. Overall, the proposed method provides a useful tool for incorporating prior information with newly collected data to improve survival prediction.

\backmatter

\section*{Acknowledgements}
The authors would like to thank Dr. Kirsten Herold at UM-SPH writing lab for her helpful suggestions on the presentation of the manuscript. The work is partially supported by  the National Institutes of Health under award numbers  R01 DK129539.

\section*{Data Availability Statement}

The SRTR risk adjustment model is available through \url{https://www.srtr.org/tools/posttransplant-outcomes/} (PSR Release Date: July 2021). The internal data used in this study is confidential and restricted to the public.





\bibliographystyle{biom} 
\bibliography{KL_bib.bib}

\section*{Supporting Information}

Web Appendices, Figures, and Tables referenced in Sections~\ref{s:method}, \ref{s:sim}, and \ref{s:real} are available with this paper at the Biometrics website on Wiley Online Library. The code for the proposed KL-based discrete failure time models are available online at: \url{https://github.com/UM-KevinHe/DiscreteKL}.

\label{lastpage}

\end{document}

%% file: preamble_ENAR2022.tex
\usepackage{bm}
\usepackage{amssymb,mathtools}
\usepackage{natbib}
\usepackage[hyphens]{url}
\usepackage[linesnumbered, ruled]{algorithm2e}
\usepackage{hyperref}
\hypersetup{
colorlinks=true,
linkcolor=red,
filecolor=magenta,
citecolor=blue,
urlcolor=blue,
}
\usepackage{cleveref}
\Crefname{assumption}{Assumption}{Assumptions}
\usepackage{setspace}
\usepackage{tikz}
\usepackage{calc} 
\tikzset{>=latex} 

\usepackage{bbm}
\usepackage{textcomp} 

\usepackage{multirow,multicol,makecell,booktabs}

\def\argmax{\operatornamewithlimits{argmax}}